\documentclass[aps, prd, tightenlines, floatfix, twocolumn, amsmath, superscriptaddress, showpacs, showkeys]{revtex4-1}
\usepackage{graphicx}
\usepackage{dcolumn}
\usepackage{bm}
\usepackage{float}
\usepackage{hyperref}
\usepackage{bigints}
\usepackage{xcolor}

\begin{document}

\title{Gamma Ray Burst GRB 221009A: two distinct hints at once at new physics}

\author{Giorgio Galanti}
\email{gam.galanti@gmail.com}
\affiliation{INAF, Istituto di Astrofisica Spaziale e Fisica Cosmica di Milano, Via Alfonso Corti 12, I -- 20133 Milano, Italy}

\author{Marco Roncadelli}
\email{marcoroncadelli@gmail.com}
\affiliation{INFN, Sezione di Pavia, Via Agostino Bassi 6, I -- 27100 Pavia, Italy}
\affiliation{INAF, Osservatorio Astronomico di Brera, Via Emilio Bianchi 46, I -- 23807 Merate, Italy}

\author{Giacomo Bonnoli}
\email{giacomo.bonnoli@inaf.it}
\affiliation{INAF, Osservatorio Astronomico di Brera, Via Emilio Bianchi 46, I -- 23807 Merate, Italy}

\author{Lara Nava}
\email{lara.nava@inaf.it}
\affiliation{INAF, Osservatorio Astronomico di Brera, Via Emilio Bianchi 46, I -- 23807 Merate, Italy}
\affiliation{INFN, Sezione di Trieste, Via Alfonso Valerio 2, I -- 34127 Trieste, Italy}

\author{Fabrizio Tavecchio}
\email{fabrizio.tavecchio@inaf.it}
\affiliation{INAF, Osservatorio Astronomico di Brera, Via Emilio Bianchi 46, I -- 23807 Merate, Italy}

\date{\today}
\begin{abstract}
The brightest ever observed gamma ray burst GRB 221009A at redshift $z = 0.151$ was detected on October 9, 2022. Its highest energy photons have been recorded by the LHAASO collaboration up to above $12 \, \rm TeV$, and one of the at ${\cal E} = 251 \, \rm TeV$ by the Carpet-2 collaboration. Very recently, the Carpet-3 collaboration has completed the data analysis, showing that the evidence of the $251 \, {\rm TeV}$ photon is quite robust. Still, according to conventional physics photons with ${\cal E} \gtrsim 10 \, \rm TeV$ cannot be observed owing to the absorption by the extragalactic background light (EBL). Previously it has been demonstrated that an axion-like particle (ALP) with allowed parameters ensures the observability of the LHAASO photons. Here we show that the Lorentz invariance violation allows the ${\cal E} = 251 \, {\rm TeV}$ (now around 300 TeV) Carpet photon to be detected.
\end{abstract}

\maketitle


\section{Introduction}

The exceptionally bright gamma ray burst GRB 221009A -- also called the brightest of all times (BOAT)~\cite{boat} -- has been detected on October 9, 2022 by the Swift observatory~\cite{swift} and by the Fermi Gamma-ray Burst Monitor (Fermi-GBM)~\cite{fermi} at redshift $z = 0.151$~\cite{red1}. Specifically, the observational situation of present interest can be summarized as follow. More than 60,000 photons have been recorded by the Water Cherenkov Detector Array (WCDA) of the LHAASO collaboration in the energy range $200 \, {\rm GeV} \leq {\cal E} \leq 7 \, {\rm TeV}$ during 3,000 \, s after the Fermi-GBM trigger (henceforth, the trigger)~\cite{LHAASOspectrum2}. In addition, 142 photon-like events have been registered by the KM2A detector -- also of the LHAASO collaboration -- in the energy range $3 \, {\rm TeV} \leq {\cal E} \leq 20 \, {\rm TeV}$ over the time span is $230 \, {\rm s} \leq t \leq 900 \, {\rm s}$ after the trigger, 9 of them with ${\cal E} \gtrsim 10 \, {\rm TeV}$~\cite{LHAASOspectrum2}. Finally, a single photon of energy ${\cal E} = 251 \, {\rm TeV}$ has been observed at $t = 4536 \, {\rm s}$ after the trigger by the Carpet-2 collaboration~\cite{carpet2}. Actually, the apparatus used by Carpet-2 consisted of an inner small area muon (ISAM) detector and an outer large area (OLAM) detector -- both of which were operative -- but the first reported result was based by the data collected by the ISAM detector alone. Quite recently, the Carpet-3 collaboration has come back to its previous observation, releasing the analysis of the data collected by the OLAM detector and only subsequently analyzed: the event at ${\cal E} = 251 \, {\rm TeV}$ has been confirmed, which has now become quite robust~\cite{carpet}. We stress that photons with ${\cal E} \gtrsim 10 \, {\rm TeV}$ cannot be observed within conventional physics (CP) because of the absorption of the extragalactic background light (EBL)~\cite{ebl}. Our aim is to show that the Lorentz Invariance Violation (LIV)~\cite{addazi} -- a potential signature of quantum gravity -- allows for the observability of the Carpet result. Recalling that an axion-like particle (ALP) with allowed parameters allows the observability of the LHAASO events, we find it extremely rewarding that a single astronomical source provides two hints a different topics belonging to the physics beyond the standard model.

\section{GRB 221009A spectrum}

We evaluate the intrinsic (de-absorbed) GRB 221009A spectrum ${\cal F}_{\rm int}$ up to 300 TeV to study the Carpet event~\cite{carpet} by starting from the LHAASO event up to $(13-18) \, \rm TeV$~\cite{LHAASOspectrum,LHAASOspectrum2}. In particular, as rule of thumb we extrapolate up to 300 TeV the spectrum reported by LHAASO in~\cite{LHAASOspectrum} obtained for energies below 10 TeV. This is justified since no cutoff is observed at higher energies~\cite{LHAASOspectrum2}. We follow the same steps reported in our Letter~\cite{gnrtb}, to which we send the reader. 

We evaluate the observed spectrum ${\cal F}_{\rm obs}$ by employing 
\begin{equation}
{\cal F}_{\rm obs}({\cal E}) = P({\cal E}; \gamma \to \gamma) {\cal F}_{\rm int}({\cal E})~, 
\label{spectrum}
\end{equation}
where $P({\cal E}; \gamma \to \gamma)$ is the photon survival probability in the different cases: $P({\cal E}; \gamma \to \gamma)=P_{\rm CP}({\cal E}; \gamma \to \gamma)$ in the CP case, $P({\cal E}; \gamma \to \gamma)=P_{\rm ALP}({\cal E}; \gamma \to \gamma)$ for the ALP case, and $P_{\rm LIV}({\cal E}; \gamma \to \gamma)$ in the case of the LIV scenario.

\section{Lorentz invariance violation}

Several attempts to describe gravity in a quantum fashion predict the violation of the Lorentz invariance above a high-energy scale ${\cal E}_{\rm LIV}$ (for a review, see~\cite{addazi}). Depending on the model parameters, LIV exhibits many different effects, altering standard physical interactions and enabling processes that would be prohibited, such as: photon decay, the vacuum Cherenkov effect, photon splitting, changes to existing thresholds in particle reactions~\cite{liberati2013}. What is important in the present context is the LIV-induced modification of the photon dispersion relation
\begin{equation}
{\cal E}^2-p^2 = - \frac{{\cal E}^{n+2}}{{\cal E}_{{\rm LIV}}^n}~,
\label{liv}
\end{equation}
with ${\cal E}$ and $p$ the photon energy and momentum, respectively. Equation~\ref{liv} produces a modification in the threshold of the $\gamma\gamma \to e^+e^-$ process: hundreds TeV photons which mainly interact with far-IR EBL and CMB photons in the standard scenario, when LIV effect is present, interact with higher energies photons, with a resulting decrease of the photon optical depth~\cite{gtl,tavLIV}.

\section{Results}

\begin{figure}
\begin{center}
\includegraphics[width=.45\textwidth]{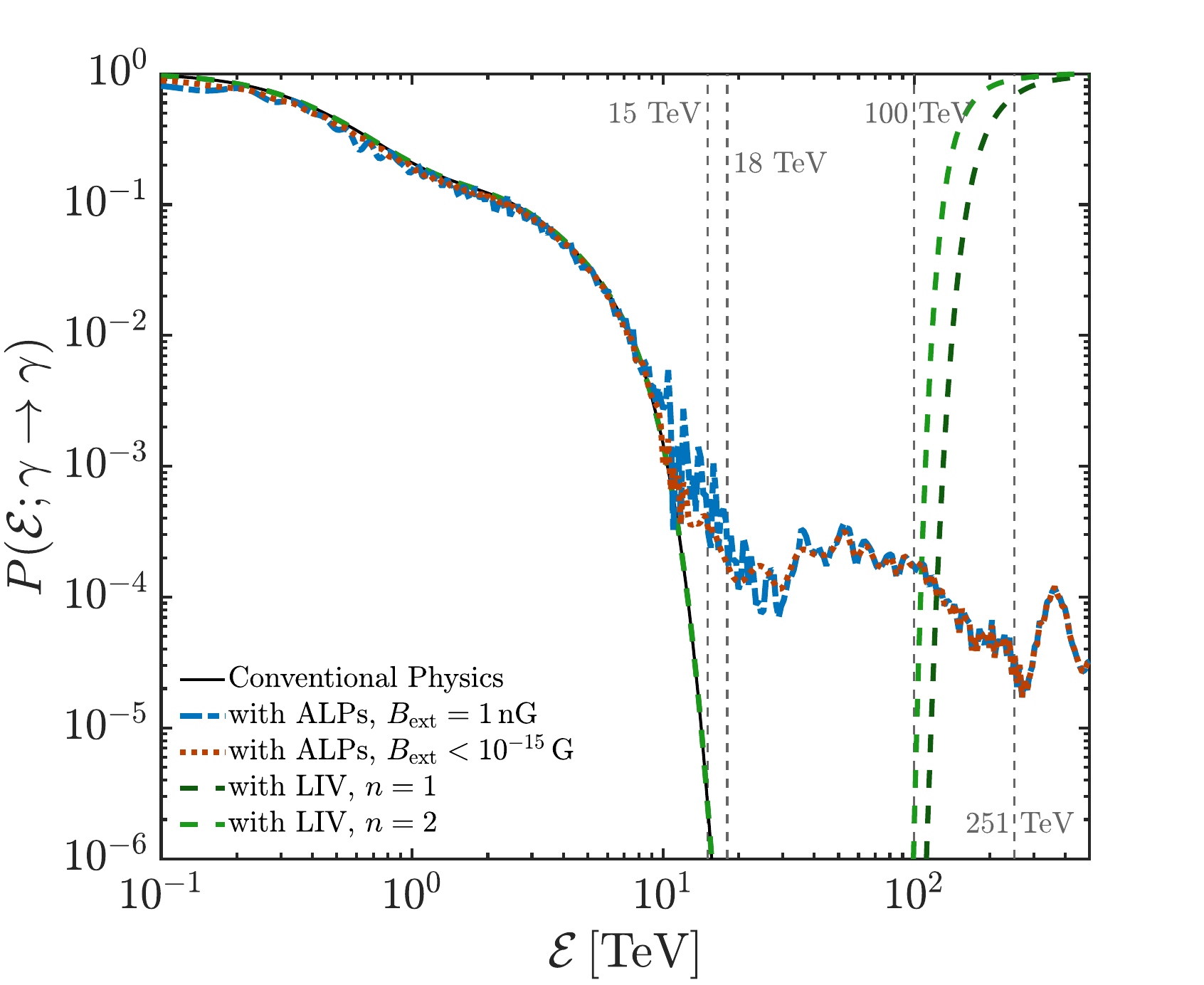}
\end{center}
\caption{\label{survProbFigStarburst} Photon survival probability $P ({\cal E}; \gamma \to \gamma)$ versus energy ${\cal E}$ in conventional physics, taking into account LIV and ALP effects.  Credit~\cite{gnrtb}.}
\end{figure}

Remaining within the current LIV bounds~\cite{LIVlim}, Fig.~\ref{survProbFigStarburst} reported in~\cite{gnrtb} shows that in both the cases of $n=1$ taking ${{\cal E}}_{{\rm LIV}, n=1} = 3 \times 10^{29} \, {\rm eV}$ and $n=2$ assuming ${{\cal E}}_{{\rm LIV}, n=2} = 5 \times 10^{21} \, {\rm eV}$, the photon survival probability when LIV effects are considered $P_{\rm LIV}({\cal E}; \gamma \to \gamma)$ approaches the value $P_{\rm LIV}({\cal E}; \gamma \to \gamma)=1$ when ${\cal E} \simeq 300 \, \rm TeV$.
In Fig.~\ref{survProbFigStarburst} the alternative ALP model for GRB 221009A is also shown, for a complete discussion and the assumed parameters see~\cite{gnrtb}.

We now want to understand how much conventional physics (CP) fails to explain the Carpet event around 300 TeV and if ALP and/or LIV models can instead justify such a detection. In particular, we evaluate how many photons can be observed by Carpet within the different scenarios: (i) within CP, (ii) with ALPs, (iii) with LIV. In order to obtain this result, we start from the LHAASO spectrum reported in~\cite{LHAASOspectrum} and discussed in Sec. II: we assume that the power-law behavior extends up to the Carpet event at 300 TeV for reference. Note that LHAASO does not observe any cutoff up to $(13-18) \, \rm TeV$~\cite{LHAASOspectrum2}. We assume an exposure time of 4000 s and conservatively a Carpet effective area of $\sim 60 \, \rm  m^2$ following the description in~\cite{carpet}. The photon count can be simply obtained by integrating the observed spectrum of Eq.~\ref{spectrum} in the different scenarios (CP, ALPs, LIV) over the energy range under consideration and reported by Carpet~\cite{carpet} reading $ 262 \, {\rm TeV} \le {\cal E} \le 343 \, {\rm TeV}$ and then multiplying by the exposure time and by the Carpet effective area (see above). In order to explain the Carpet observation  we must obtain a photon count $N$ around 1. Our results are as follows. We obtain a photon count in the different models equals to: (i) $N=1.8 \times 10^{-97}$ in CP, (ii) $N=2.9 \times 10^{-6}$ with ALPs, (iii) $N=0.1$ with LIV in the case of both $n=1$ and $n=2$.

\section{Discussion and Conclusions}
As we can observe from our results CP cannot explain the Carpet event around 300 TeV~\cite{carpet} but even the ALP model which explains the LHAASO event at $(13-18) \, \rm TeV$~\cite{LHAASOspectrum2} can hardly explain this event. Instead, the considered LIV models reasonably provide an explanation for this event but are unable to explain the LHAASO event. Therefore, GRB 221009A seems to provide two hints at new physics: (i) at ALPs concerning the LHAASO event at $(13-18) \, \rm TeV$~\cite{LHAASOspectrum2} as shown in~\cite{gnrtb} and (ii) at LIV regarding the Carpet event around 300 TeV~\cite{carpet}.


\begin{thebibliography}{9} 



\bibitem{boat} E. Burs {\it et al.}, Astrophys. J. Lett. {\bf 946}, L21 (2023). 

\bibitem{swift} M. A. Williams {\it et al.}, Astrophys. J. Lett. {\bf 946}, L24 (2023). 

\bibitem{fermi} S. Lesage {\it et al.}, Astrophys. J. Lett. {\bf 952}, L42 (2023). 

\bibitem{red1} A. de Ugarte Postigo {\it et al.}, GRB 221009A: GRB coordinates network {\bf 32648}, 1 (2022).






\bibitem{LHAASOspectrum2} LHAASO Collaboration, Science Advances {\bf 9}, eadj2778 (2023).



\bibitem{carpet2} D.Dzhappuev {\it et al.} [Carpet-2 collaboration], The Astronomer's Telegram {\bf 15669} (2022).







\bibitem{carpet} Carpet-2 Collaboration, arXiv:2502.02425 (2025).


\bibitem{ebl} A. Saldana-Lopez {\it et al.}, Mon. Not. R. Astron. Soc. {\bf 507}, 5144 (2021).







\bibitem{addazi} A. Addazi {\it et al.}, Progr. in Part. and Nucl. Phys. {\bf 125}, 103948 (2022).


\bibitem{LHAASOspectrum} LHAASO Collaboration, Science {\bf 380}, 1390 (2023).

\bibitem{gnrtb} G. Galanti, L. Nava, M. Roncadelli, F. Tavecchio and G. Bonnoli, Phys. Rev. Lett. {\bf 131}, 251001 (2023).

\bibitem{liberati2013} S. Liberati, Class. Quant. Grav. {\bf 30}, 133001 (2013).

\bibitem{gtl} G. Galanti, F. Tavecchio and M. Landoni, Mon. Not. R. Astron. Soc. {\bf 491}, 5268 (2020).

\bibitem{tavLIV} F. Tavecchio and G. Bonnoli, Astron. Astrophys. {\bf 585}, A25 (2016).


\bibitem{LIVlim}  R. G. Lang, H. Mart\'inez-Huerta, and V. de Souza, Phys. Rev. D {\bf 99}, 043015 (2019).




\end{thebibliography}
\end{document}